\documentclass[galaxies,conferencereport,accept,moreauthors,10pt,a4paper]{mdpi}
\usepackage{gensymb}
\usepackage{booktabs}
\usepackage{multirow}
\usepackage{soul}
\usepackage{microtype}

\firstpage{1}
\articlenumber{x}
\doinum{10.3390/------}
\pubvolume{xx}
\pubyear{2017}
\copyrightyear{2017}
\externaleditor{Academic Editor: Emilio Elizalde}
\history{Received: 18 July 2017; Accepted: 24 August 2017; Published: date}

\usepackage{amssymb}

\Title{The HI Distribution Observed toward a Halo Region of the Milky Way}

\Author{{Ericson L\'opez }*, Jairo Armijos , Mario Llerena  and Franklin Ald\'as }
\AuthorNames{L\'opez Ericson, Armijos Jairo, Llerena Mario and Ald\'as Franklin.}

\address[1]{%
Observatorio Astron\'omico de Quito, Escuela Polit\'ecnica Nacional, Av. Gran Colombia S/N, {Quito 170403}, Ecuador; \\{jairo.armijos@epn.edu.ec (J.A.); mario.llerena01@epn.edu.ec (M.L.); franklin.aldas@epn.edu.ec (F.A.)}}

\corres{\hangafter=1 \hangindent=1.05em \hspace{-0.82em}Correspondence: {ericsson.lopez@epn.edu.ec}; Tel.: +593-22570765}

\abstract{We use observations of the neutral atomic hydrogen (HI) 21-cm emission line to study the spatial distribution of the HI gas in a 80$\degree\times~$90$\degree$ region of the Galaxy halo. The HI column densities in the range of \mbox{3--11$\times$10$^{20}$ cm$^{-2}$} have been estimated for some of the studied regions. In our map---obtained with a spectral sensitivity of $\sim$2 K---we do not detect any HI 21-cm emission line {above 2$\sigma$} at Galactic latitudes higher than $\sim${46}$\degree$. {This report summarizes our contribution presented at the conference on the origin and evolution of barionic Galaxy halos}.}


\conferencetitle{On the Origin (and Evolution) of Baryonic Galaxy Halos}

\begin{document}



\section{Introduction}
Neutral atomic hydrogen (HI) is the most abundant element in the interstellar medium, and its \mbox{21-cm} emission line is a powerful tool to trace the structure and dynamics of the Milky Way Galaxy (Kalverla and J\"urgen \cite{Kalberla}). The HI gas in our own galaxy has a {two-component} structure; one is  composed of cold neutral gas with temperatures {$\lesssim$}300 K and the other by warm neutral gas with temperatures {$\gtrsim$}5000 K (Kalverla and J\"urgen \cite{Kalberla}). The HI disk of the Milky Way together with its spiral arms extend up to a radius of $\sim$35 kpc (Kalverla and J\"urgen \cite{Kalberla}). There is also a galaxy halo composed of HI gas with densities of 10$^{-3}$ cm$^{-3}$ that extends in vertical height up to a distance of $\sim$4 kpc, and radially HI gas is detected in the outskirts ({$\gtrsim$}35 kpc) of the Milky Way (Kalverla and J\"urgen \cite{Kalberla}). \linebreak All these important features have been {discovered} mainly thanks to studies carried out using the HI 21-cm emission line.

{The HI column density (N$_{\rm{HI}}$) has been estimated across the galaxy by Dickey and Lockman \cite{Dickey}, who estimated the highest and lowest N$_{\rm{HI}}$ values of 2.6 $\times$ 10$^{22}$ at ($l$,$b$) = (339$\degree$,0$\degree$) and of 4\mbox{.4 $\times$ 10$^{19}$ cm$^{-2}$} at ($l$,$b$) = (152$\degree$,62$\degree$), respectively.}
A recent HI 4$\pi$ survey has been used to obtain all-sky column density maps of HI for the Milky Way (Ben Bekhti et al. \cite{Ben_Bekhti}). The data of this survey have a spectral sensitivity of 43 mK and a spatial resolution of 16.2 arc-min.

In this work, we aim to map a defined region {(chosen randomly)} of the Milky Way ($87\degree  \leq  l \leq 180  \degree $ and 13$\degree$ $\leq$ $b$ $\leq$ 180 $\degree$) at the HI 21-cm emission line, { which can help us to identify a disk-halo transition region}. {The data employed in our study have been obtained using} the 2.3 m SALSA radio {telescopes}\footnote{\url{http://vale.oso.chalmers.se/salsa/}} of the Onsala Space Observatory located in Sweden, which are operating at a wavelength of 21 cm. We also aim to check whether our N$_{\rm{HI}}$ estimates derived for our selected regions are in good agreement with previous values derived by other authors (Dickey and Lockman \cite{Dickey}, \linebreak Ben Bekhti et al. \cite{Ben_Bekhti}). Even working with a small telescope mainly used for student experiences, we expect to achieve acceptable results in comparison with those observations obtained from larger telescopes like Parkes and Dwingeloo.


\section{Observations and Data Reduction}\label{Obs}

\begin{figure}[H]
\centering
\includegraphics[width=0.35\textwidth,trim={0cm 0cm 0cm 0cm},clip ]{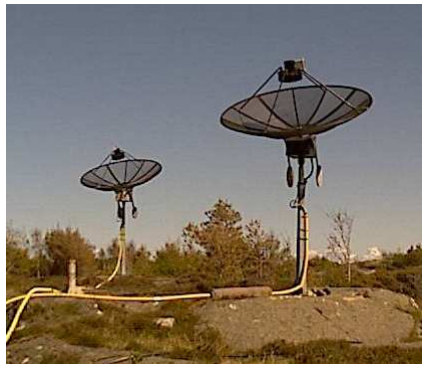}
\caption{The 2.3 m radio telescopes of the Onsala Space Observatory.}\label{fig1}
\end{figure}

As mentioned previously, we used two small radio telescopes (see Figure~\ref{fig1}) of the Onsala Space Observatory to carry out our observations. The observations were performed in September 2016. The telescopes provide a spatial resolution of 6$\degree$ at 21 cm, and their receivers have a spectral resolution of 7.8 KHz per channel and 2 MHz bandwidth \cite{Salsa_manual}. Using both telescopes, we mapped a region between galactic longitudes 87$\degree$ and 180$\degree$ and galactic latitudes 13$\degree$ and 85$\degree$, in steps of 5.5$\degree$.

The data processing and analysis was done using the SalsaSpectrum {software}\footnote{ \url{http://vale.oso.chalmers.se/salsa/software}}. Unfortunately, the data obtained with the 2.3 m radio telescopes are not flux calibrated. To carry out this process, we observed several positions toward the galactic plane. Then, the obtained HI 21-cm spectra were compared with calibrated spectra of the same galactic positions previously observed (Higgs and Tapping \cite{Higgs}), obtaining in this way a flux calibration factor. The calibration observations were carried out toward four positions: $l$ = 74$\degree$, $b$ = 1$\degree$; $l$ = 119$\degree$, $b$ = $-$1$\degree$; $l$ = 131.2$\degree$, $b$ = 1$\degree$; and $l$ = 140$\degree$, \mbox{$b$ = $-$3$\degree$}. The intensity of our calibrated spectra is given in brightness temperature (T$_B$*). { We noted that the HI 21-cm line intensities---observed toward the calibration positions---did not change by more than 20\% as the integration time varied from 20 s to 160 s; therefore, our spectra are affected by 20\% uncertainties in the intensity}.
We used an integration time of 20 s to observe the calibration positions and the target regions of the Milky Way.
\vspace{6pt}


\section{Results}

\begin{figure}[H]
\centering
\includegraphics[width=1\linewidth,trim={0cm 14cm 0cm 0cm},clip ]{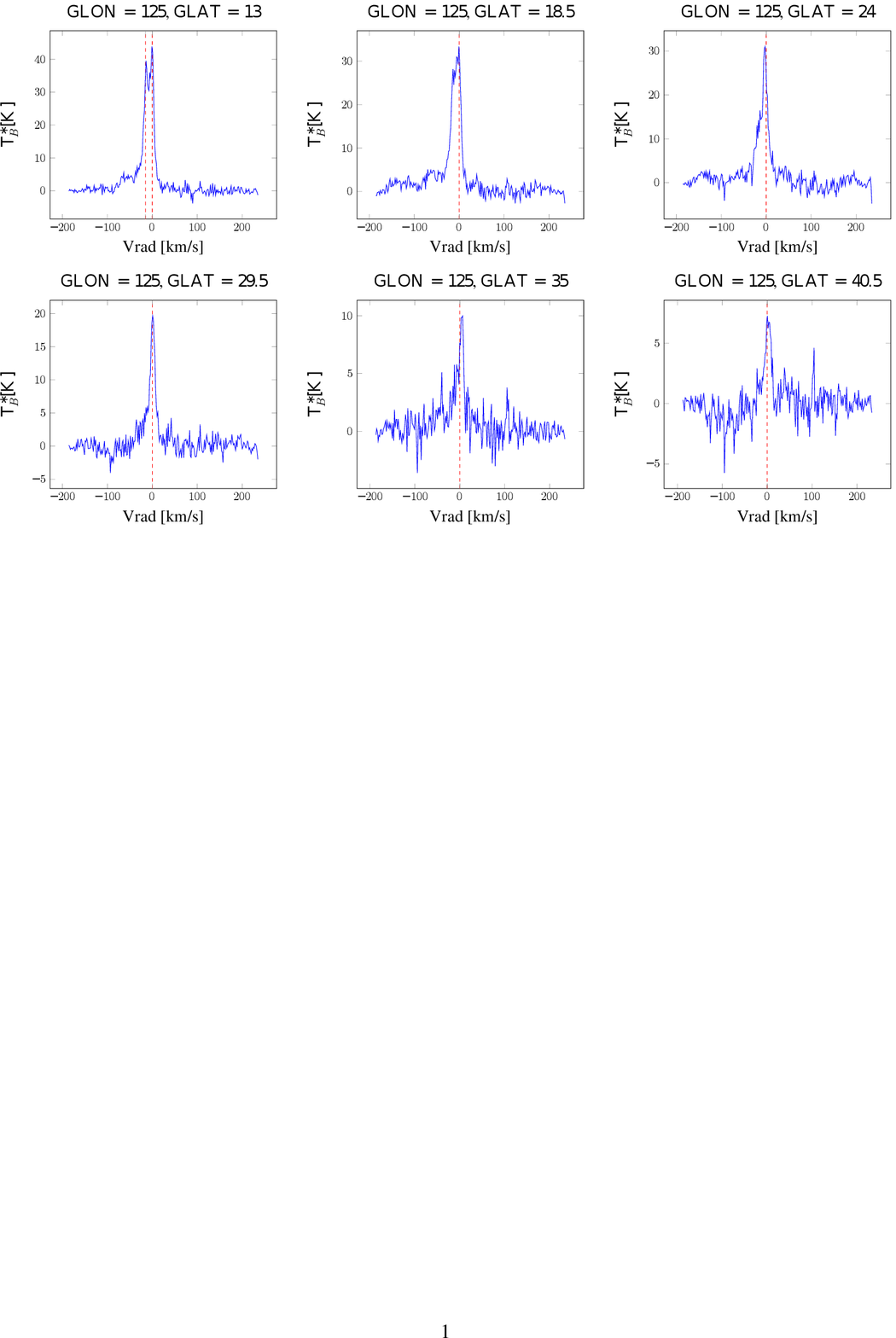}
\caption{{Neutral atomic hydrogen (HI) 21-cm}  emission line observed toward six positions in the Milky Way. The dashed red line indicates the 0 km s$^{-1}$ velocity in all of the panels. In the upper-left panel, the velocity of $-20$ km s$^{-1}$---corresponding to another gas component with a radial velocity different than 0 km s$^{-1}$---is also {indicated.}
}\label{fig2}
\end{figure}

\begin{figure}[H]
\centering
\includegraphics[width=0.6\linewidth,trim={0cm 0cm 0cm 0cm},clip]{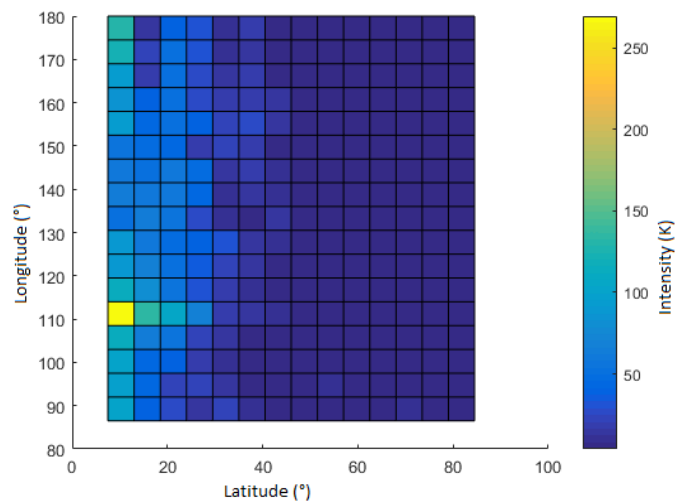}
\caption{Peak-intensity map of HI 21-cm lines covering an 80$\degree$ $\times$ 90$\degree$ area of the Milky Way.}\label{fig3}
\end{figure}

As we expected, the strongest HI 21-cm lines were detected at the lowest galactic latitudes ($\sim$8$\degree$) covered in our study. In Figure~\ref{fig2} we show a sample of HI 21-cm emission profiles observed toward six selected positions in the Milky Way. As seen in this figure, the 21-cm lines show central line velocities of $\sim$0 km s$^{-1}$, that correspond at low latitudes to our own galactic arm and at high latitudes to the galactic halo. The spectrum observed at $l$ = 125$\degree$ and $b$ = 13$\degree$ shows the HI emission line clearly tracing two gas components with velocities of 0 and $-$20 km s$^{-1}$. The gas with negative velocities is likely associated with another spiral arm of the galaxy.
{We also show a peak-intensity map of the HI lines in Figure~\ref{fig3}. We did not detect a HI 21-cm  emission line above 2$\sigma$ ($\sigma$ $\approx$ 2 K) at galactic latitudes higher than $\sim${46}$\degree$}.
{For these regions, in our map we used 2$\sigma$ limits in the peak-intensity.}
This fact defines the lower boundary of the galactic halo, so we use this criterion based on the lack of a HI 21-cm emission line { above 2$\sigma$} to define the start of the galactic halo.
{Nevertheless, another criteria may be used to define the start of the galactic halo; for example, a threshold in the decreasing number of Milky Way HI clouds as a function of the galactocentric radius (Ford et al. \citep{Ford10}).}

On the other hand, we used the SalsaSpectrum software {to subtract the underlying continuum emission} and fit Gaussians to the H 21-cm lines. Then, we were able to estimate the integrated line intensities ($\int T_b dV$), which are listed in Table \ref{tab1} for the galactic positions indicated in Figure~\ref{fig2}.
In this figure, the last two HI 21-cm line profiles ({GLAT}: 35 $\degree$ and 40.5 $\degree$) show the lowest signal-to-noise ratio, so they were not considered.
The remaining profiles were used to derive the hydrogen column density (N$_{\rm{HI}}$), which is proportional to $\int T_b dV$ (assuming an optically thin medium), and it can be estimated following the expression given by Dickey and Lockman \cite{Dickey}:

\begin{equation}\label{eq1}
\left(\dfrac{N_H}{\text{cm$^{-2}$}}\right)= 1.82\times10^{18} \int  T_b dV,
\end{equation}

Using this expression, we estimated the hydrogen column density for our halo sub-regions confined between 13\degree ~and 30\degree of galactic latitude. The obtained values are given in Table \ref{tab1}.

\begin{table}[H]
\centering
\caption{Parameters derived for four positions of the Milky Way}\label{tab1}
\begin{tabular}{cccc}
\toprule
\boldmath{$l$} & \boldmath{$b$}  &  \boldmath{$\int T_b dV$}   &  \bf{N$_{\rm{H}}$} \\
 \bf{Degrees} &  \bf{Degrees} & \ \bf{(K km s$^{-1}$)} &  \bf{($\times$10$^{20}$ cm$^{-2}$)}\\
\midrule
125 & 13.0 & 623.5 & 11.4\\
125 & 18.5 & 514.7 &  9.4\\
125 & 24.0 & 391.7 &  7.1\\
125 & 29.5 & 193.4 &  3.5\\
\bottomrule
\end{tabular}
\end{table}
\vspace{-12pt}

\section{Conclusions and Future Work}

Using data obtained from the small SALSA telescope, we were able to obtain a peak-intensity map of the HI 21-cm emission line for a 80$\degree$ $\times$ 90$\degree$ region of the Milky Way. We did not detect any emission line { above 2$\sigma$} at Galactic latitudes higher than $\sim${ 46}$\degree$, so we have considered this latitude to define the galactic halo lower boundary.

 {We have also estimated atomic hydrogen column density N$_{\rm{HI}}$ values, which fall in the range of 3--11 $\times$ 10$^{20}$ cm$^{-2}$, for our four randomly-chosen positions of the Milky Way, under the assumption of an optically thin medium}.

{ Despite the low spatial resolution of our observations, the derived N$_{\rm{HI}}$ values are consistent with the values of $\sim$10$^{20}$--10$^{21}$ cm$^{-3}$, provided for the same regions studied in our paper with a much higher spatial resolution instrument (Ben Bekhti et al. \cite{Ben_Bekhti}). In the same context, our results are also in good agreement with the N$_{\rm{HI}}$ range of 4.4 $\times$ 10$^{19}$ $-2.6$ $\times$ 10$^{22}$ cm$^{-2}$ obtained by Dickey and Lockman \cite{Dickey} in their study of the HI gas across the galaxy.

{Regarding our future work, instead of continuing the study of the large-scale distribution of HI gas in the Milky Way, we are interested in the study of the physical and dynamical properties of galactic high-velocity clouds. For that, we plan to apply for observing time in large radio telescopes such  IRAM, Parkes, among others---instruments with much better spatial resolutions and data quality}.

\vspace{6pt}



\vspace{6pt}
\authorcontributions{All authors carried out the observations and performed the data reduction and analysis. E. L\'opez and J. Armijos wrote the manuscript.}

\conflictsofinterest{The authors declare no conflict of interest.}

\end{document}